\documentclass[aps,prd,twocolumn,groupedaddress,letterpaper,nofootinbib]{revtex4}
\usepackage{graphicx}

\newcommand{\Msun}{\ensuremath{M_{\odot}}}
\newcommand{\rholambda}{\rho_{\Lambda}}
\newcommand{\dFdM}{\frac{\mathrm{d}F}{\mathrm{d}M}}
\begin{document}


\title{Effects of Inhomogeneity
  on the Causal Entropic prediction of $\Lambda$}


\author{Daniel Phillips}
\affiliation{Department of Physics, UC Davis}
\author{Andreas Albrecht}
\affiliation{Department of Physics, UC Davis}


\date{\today}

\begin{abstract}
The causal entropic principle aims to predict the
unexpectedly small value of the cosmological constant $\Lambda$ using
a weighting by entropy increase on causal diamonds. The original work
assumed a purely isotropic and homogeneous cosmology. But even the
level of inhomogeneity observed in our universe forces reconsideration
of certain arguments about entropy production. In particular, we must
consider an {\em ensemble} of causal diamonds associated with each background cosmology
and we can no longer immediately discard entropy production in the far
future of the universe. Depending on our choices for a probability
measure and our treatment of black hole evaporation, the prediction
for $\Lambda$ may be left intact or dramatically altered.
\end{abstract}


\maketitle



\section{Introduction}
A broad line of argument intended to resolve or ameliorate the notorious problem of
the apparent smallness of the cosmological constant ($\rholambda
\approx 1.25 \times 10^{-123}$ in Planck units) is to reject the notion of
a fundamental value for $\Lambda$ altogether. In this approach,
well-known from the string theory landscape as well as other
``multiverse'' notions, the problem is transformed to the search for a
selection principle that may explain why a value as small as observed
is probable. In order to make this formulation two broad decisions
must be made, both of which can be controversial: the choice of selection
principle, and the probability measure.  The first tends to be controversial
because a choice of selection principle is a choice about how to
categorize our imagined experimental sample of universes in which
measurements occur, and thus leads to difficult questions about
observers. The choice of probability measure has its own well-known
difficulties relating to defining probabilities across different
infinite spaces.  Different choices for either selection principle or
probability measure can lead to wildly different probability
predictions, easily changing a prediction of likelihood to an
exponentially disfavored one.

\begin{figure}[h]
  \includegraphics[height=.22\textheight]{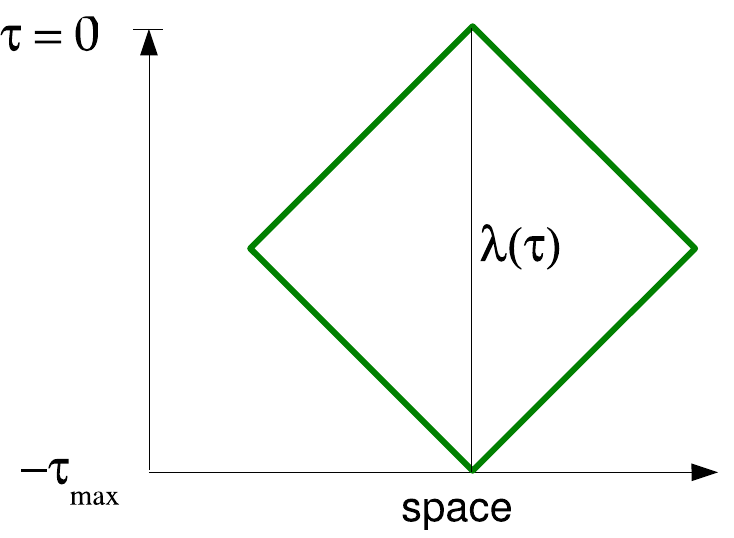}
 \caption{A causal diamond (depicted schematically here) is the region
 which can causally impact {\em and} be causally impacted by a
 worldline $\lambda(\tau)$.  The finite entropy produced in the
 resulting spacetime volume is used in the causal entropic
 principle as a cosmological weighting factor} 
 \label{diamond}
\end{figure}

Bousso {\it et al.} \cite{Bousso} suggested a novel combined approach, the
so-called ``causal entropic principle'' (CEP).  For flat universes
with a positive fundamental cosmological 
constant, one can define the \emph{causal diamond} for a particular
world line $\lambda(\tau)$ as the 
intersection of interiors of the future cone at earliest times and the
past cone at late times\footnote{The CEP has been
  extended to provide predictions of curvature and other cosmological features  in \cite{Bozekxxx, BoussoLandscape}}. 
 The resulting region is finite in comoving
volume in this flat positive-lambda Friedmann-Robertson-Walker (FRW) universe, and 
diamond-shaped when drawn in comoving coordinates and conformal time. [See
Fig. \ref{diamond}].  If we restrict our probability measure to the
finite interior of this diamond, we can avoid the difficulty in
defining a probability measure on infinite spaces.  Moreover, the
proposed selection principle is a simple weighting proportional to the
entropy production $\Delta S$ occurring within the causal
diamond. Loosely one may interpret this as an assumption that the
number of observers is proportional to the entropy increase within the
causal diamond, but in the spirit of \cite{Bousso} we may simply take
this weight as a hypothesis and remark that $\Delta S$ has several
advantages over some other weightings: 1) It is hearteningly generic,
allowing at least the theoretical possibility of application to
universes with much different low-energy physics from ours. 2) It
seems less contrived than typical ``anthropic'' reasoning; though we
may contemplate observers in a universe with no galaxies, it is difficult
to imagine them without significant entropy increase. 3) As shown in
\cite{Bousso}, it can actually reproduce and improve upon previous
anthropic results. This work has since been taken in a number of interesting directions \cite{Freivogel:2011eg,Bousso:2010im,Bousso:2010zi}.

Even after accepting the program to calculate likelihoods of physical
 parameters from some a priori theoretical distribution and after
 fixing a probability measure, a full calculation of the probability
distribution for $\Lambda$ is a formidable task.  In an ideal case we
 would have a background theory giving us some set of cosmological
 parameters and their prior distribution.  We would then allow all
 parameters to vary and make a prediction for $\Lambda$ by
 marginalizing over the other parameters, in a scheme such as that in
 \cite{Tegmark}.  As a first step, Bousso {\it et al.} \cite{Bousso} follow
 the usual simplification of holding all other physical parameters
 fixed while modifying only the positive value of $\Lambda$ in a flat
 FRW universe. Other work has discussed aspects of the CEP
 for $\Lambda \leq 0$ \cite{Mersini-Houghton, Salem}, but in this work
 we keep the same $\Lambda > 0$ assumption used in the first papers
 on this subject.

A common drawback of varying only $\Lambda$ in such an approach is the
possibility that variation in other parameters could significantly
affect the prediction for $\Lambda$ itself.  The classic instance is
that Weinberg's prediction of $\rholambda < 10^{-121}$\cite{Weinberg}
under the selection principle that galaxies must form is softened by
allowing the density contrast $Q$ or the baryon-to-photon ratio to
increase from that observed in our universe.  Greater early
anisotropies or matter densities and reduced radiation pressure could
allow structure to form earlier and thus significantly push up the
allowable value of $\Lambda$ \cite{Feldstein, Garriga}.   Cline {\it
  et al.} \cite {Cline} have 
shown that the entropic approach, at least for $\Lambda$, is resilient
when varying $Q$.  More recently it has been shown that allowing the
curvature of the universe to vary along with $\Lambda$ can
dramatically change the CEP predictions for $\Lambda$, depending on
exactly what priors one take on the cosmic
curvature\cite{Bozekxxx}. Other authors have suggested potential
limitations of the CEP along with related approaches \cite{Mersini-Houghton, Maor}.

This paper examines a different simplification that has been made so
far in all work on the CEP: that of an isotropic, homogeneous
universe.  Our own universe's small primordial fluctuations allow us
to make these approximations to great effect for the overall evolution
of the universe.  But as time progresses we know that structure
formation proceeds apace and entropy production, if it is associated
with structure, becomes less spatially homogeneous. Because the causal
entropic approach examines only the causal region surrounding a
particular world line, we must try to formulate how departures from
homogeneity may affect the entropic weight, and whether those
variations can affect the prediction for $\Lambda$. (In the process we
estimate entropy production well into the era of cosmological constant
domination, but we note that our approach is not directly related to the
arguments in \cite{Mersini-Houghton} about the cosmological heat death of observers.)

In Sec. II we briefly review the CEP method. In Sec. III we
discuss the resulting prediction for $\Lambda$ and comment on the
increasing inhomogeneity of entropy produced at late times,
illustrated by black hole evaporation. In Sec. IV we describe the
necessity of replacing a single causal diamond with an ensemble
  representing the diverse possible behaviors of worldlines due to the
  inhomogeneities, even when the background cosmology is fixed.    
Sec. V discusses the nature of
long-term entropy sources that might compete with stellar entropy
production for causal diamonds containing collapsed structures.  In
Sec. VI we discuss effects on the predicted probability
distribution for $\rholambda$, and in section VII we summarize our
conclusions. Throughout we use Planck units with $\hbar = c = G = 1$.

\section{The Causal Entropic Prediction for $\Lambda$}
Simply stated, the CEP \cite{Bousso}
assumes that the probabilistic weighting for 
cosmological parameters is proportional to the increase in entropy
$\Delta S$ within a causal diamond associated with that
cosmology.  Given a multiverse populated with different cosmologies,
the CEP thus becomes a tool to calculate probability distributions for
measurements of the cosmological parameters themselves. 
Although in
principle one could ask the CEP to thus give predictions for a greater
range of cosmological parameters (see for example \cite{Bozekxxx, BoussoLandscape}), following Bousso {\it et al.} we will 
leave all cosmological parameters fixed at their observed values
except the cosmological constant $\Lambda$.

The causal diamond is defined as the volume contained within the
future cone of an early event (taken to be reheating following
inflation) as well as within the past cone of a late event on the same
world line. The causal diamond is thus the region of space in full
causal contact with a particular world line. Following the original
argument we will also restrict ourselves to purely positive $\Lambda$,
so that all cosmologies will eventually be dominated by the
cosmological constant. In every case a de Sitter horizon will thus
form and define the past light cone for the causal diamond.

The CEP choice to restrict entropy increase to that within a causal
diamond originated from a holography argument: the universe simply
does not consist of a region larger than a single causal diamond. We
will not try and argue the pros and cons of this point here, but
simply take this restriction as one of our input assumptions. 
There is however an important extra step
which we will talk about in greater detail. If an entire
cosmology is represented only by a single causal diamond, we need some
way to choose this causal diamond, or equivalently define a particular
world line associated with a particular set of cosmological
parameters. There is no difficulty doing so in a homogeneous,
isotropic universe, as all causal diamonds are identical. Such is
clearly not the case for an inhomogeneous universe, and we will thus
introduce the statistical notion of an ensemble of causal diamonds
associated with a particular cosmology. It should be emphasized that
this complication is required even with a very strictly holographic
interpretation of the causal diamond.

Black holes immediately come to mind in calculations of cosmological
entropy. The entropy associated with the formation of a black hole
horizon is explicitly excluded in the CEP, as is de Sitter horizon
entropy. This exclusion is important as a single supermassive $(10^7
\Msun)$ black hole can have an entropy of $10^{91}$\cite{Frampton},
exceeding all other nonhorizon entropy sources.  Reference \cite{Bousso} acknowledged the possibility of including the black hole entropy upon formation; in our universe, the era of peak black hole production approximately overlaps that of peak stellar entropy production, so there would probably be only a modest change in preferred values of $\Lambda$ compared to \cite{Bousso}. However, including this massively larger black hole entropy during the star-forming epoch would overwhelm and make moot the very long-term entropy sources we discuss in this paper. For the rest of this paper we will hold to the original convention of the CEP and exclude entropy associated directly with the formation of a black hole horizon.

One might object, as
noted in \cite{Bousso}, that this black hole entropy cannot be hidden forever, as
on very long time scales the black hole will evaporate and return its
entropy to the rest of the causal diamond. Bousso {\it et al.}
\cite{Bousso} argued that a typical late-time causal diamond is empty
and thus we may discount this entropy. We will examine this issue systematically in this paper.

It is also important to note that the weighting $w(\rholambda) \propto
\Delta S$ includes only entropy increase occurring within the causal
diamond.  Therefore various processes which one might imagine to be
strong contributors to entropy increase turn out not to be
significant.  For example, cosmic microwave background (CMB) 
photons represent a large amount of
current entropy, but not of entropy increase within the causal diamond
surrounding our world line.  The causal diamond at recombination
enclosed a much smaller amount of matter (and photons) than a Hubble
radius today does, so most CMB photons within our horizon must have
entered through the bottom cone of the causal diamond; these photons
do not contribute to $\Delta S$. Other events in the early universe
such as nucleosynthesis likewise contribute little to this measure of
$\Delta S$ owing to the small size of the causal diamond. So Bousso
{\it et al.} \cite{Bousso} restricted themselves to processes active
during the era of relatively large comoving scale for the causal
diamond.  One of the purposes of this paper is to examine whether the
very long times available for entropy production in the future of a
$\Lambda$-dominated universe can compensate for the small volume of
matter in causal contact with an observer once a de Sitter horizon
forms.

Varying the cosmological constant directly affects the size of the
causal diamond, with the comoving 4-volume contained proportional to
$\Lambda^{-1}$. Therefore even before accounting for the effects of
entropy production, the CEP rewards smaller values of $\Lambda$
with greater weight, owing to their larger causal diamonds, at least
measured in comoving volume. If we found entropy production to be dominated
by a process producing a constant entropy rate per comoving volume, such a
process would translate a flat prior distribution of $\rholambda$
$$\frac{dp}{d\rholambda} = \mathrm{const}$$
into a flat distribution in $\mathrm{log}(\rholambda)$
$$\frac{dp}{d\rholambda} \propto w(\rholambda) = \rholambda^{-1}$$
$$\frac{dp}{dlog(\rholambda)} \propto w(\rholambda)\rholambda = \mathrm{const}$$
The reduction from a flat distribution to one flat in log space is an
indication of how much work the causal diamond portion of the CEP is
doing on its own. For realistic entropy sources, the total entropy production
(and thus probabilistic weight) was calculated via
$$w(\rholambda) \propto \Delta S (\rholambda) = \int_0^{\infty} \mathrm{dt}
V_c(\rholambda,t)\frac{\partial^2 S(\rholambda,t)}{\partial{V_c}
  \partial{t}}$$
Here $V_c$ is the directly calculable comoving 3-volume of
the causal diamond as a function of $\Lambda$ and $t$ and $\partial{\dot{S}}/\partial{V_c}$ is the
entropy production rate per comoving volume.
 
During the current cosmological era for cosmologies similar to ours,
calculations in \cite{Bousso} revealed stars to be the greatest contributor to $\Delta
S$ due to photons absorbed and reemitted by cool dust.  This large
contribution may be seen from estimating entropy increase for a
process by $\Delta S = \frac{\Delta E}{T}$ where $\Delta E$ is the
energy released and $T$ is the typical temperature ($k_B = 1$).  For
stars, typical energies released in fusion are about 7 MeV/nucleon.
While typical stars produce visible light with an effective $T \sim
eV$, perhaps half of the photons are absorbed and re-emitted by cool
dust with a $T \sim 20 meV$.  It is the combination of high energy per
nucleon, ubiquity of stellar burning, and the low effective
temperature of much of the reprocessed starlight which gives stellar
entropy the edge over other processes.

\section{Comoving volume of universe}
\begin{figure}[h]
  \includegraphics[height=.25\textheight]{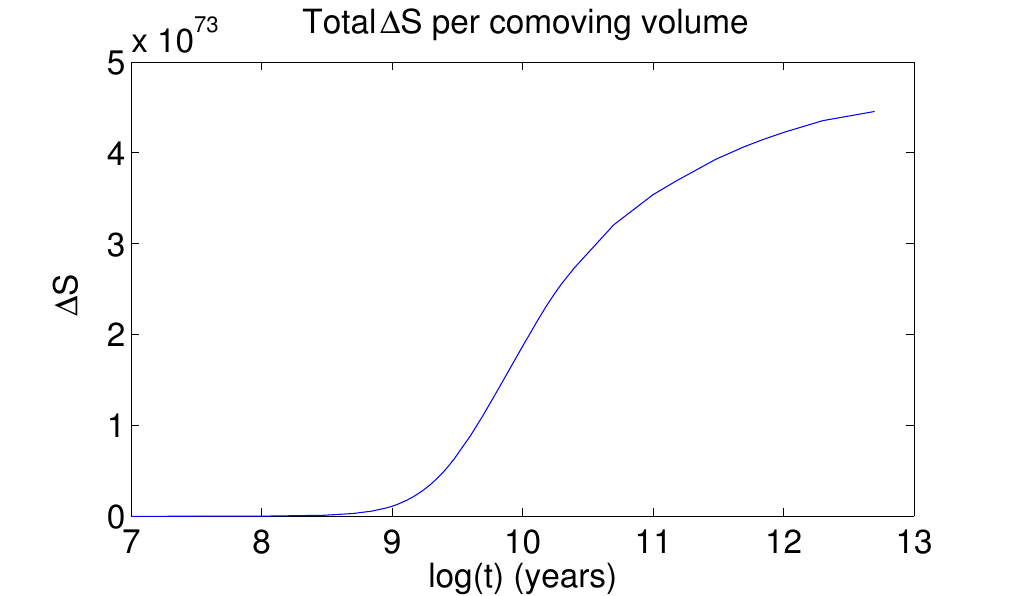}
 \caption{Integrated stellar entropy production per comoving $Mpc^3$, 
 calculated using the Nagamine {\it et al.} star formation model \cite{Nagamine} considered
 in \cite{Bousso}. The
 long tail is produced by low-mass white dwarfs with lifetimes up to $10^{13}$
 years, but by $10^{10}$ years we have already seen a large fraction of stellar entropy production.}
 \label{entpercomoving}
\end{figure}

Stellar entropy production
per comoving volume reaches a maximum of $2.7 \times 10^{63}/Mpc^3/yr$
comoving, as shown in Fig. \ref{entpercomoving}.  The causal diamond
gets as large as $\sim 10^{13} Mpc^3$. With $\sim 10^{10}$ years that gives
an integrated stellar entropy production $\Delta S_0 \approx 10^{86}$ reached
by about 10 x $10^9$ years.  

\begin{figure}[h]
  \includegraphics[height=.4\textheight]{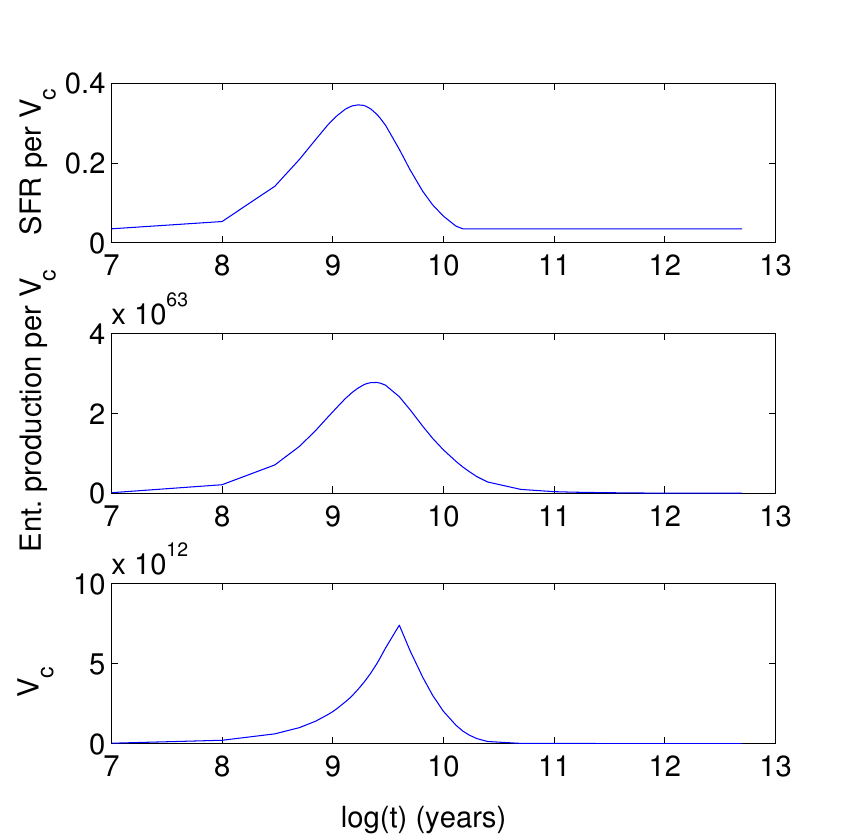}
 \caption{A peak in star formation (top plot) is followed by a peak in
 entropy production (middle plot) per comoving volume.  In our
 universe the peak in comoving 3-volume of the causal diamond (bottom
 plot) is near
 the time of maximal stellar entropy production per $V_c$. The
 3-volume $V_c(t)$ of the causal diamond is determined by $\Lambda$;
 a much earlier peak (larger $\Lambda$) would not allow the diamond to
 capture as much entropy production.  Universes with smaller
 $\Lambda$ would give larger causal diamonds in late times, but would
 capture little more stellar entropy production, and are less likely
 owing to our flat prior. All plots are assuming a homogeneous universe.}
 \label{ent_and_vol}
\end{figure}

Under the CEP with stars as the major source of entropy production, one
obtains a weighting and hence a predicted probability distribution for
$\rholambda$ [Fig. \ref{ent_and_vol}].  With several different star formation models
\cite{Bousso,Cline,BoussoStarformation} the predicted 1-$\sigma$ probability band
of roughly $10^{-124} \lesssim \rholambda \lesssim 10^{-122}$ easily
contains our universe's observed value. 

A comoving volume of a particular scale contains a fixed amount of
matter so long as the universe is homogeneous over the scale of
consideration.  But a flat universe with a cosmological constant will
form a horizon of fixed physical size. Eventually the comoving
radius corresponding to the horizon length will drop below the scale
of matter inhomogeneity. In physical terms, for a world line near a
gravitationally collapsed halo, the amount of mass enclosed by a
causal diamond will eventually approximate a constant value, rather
than exponentially emptying out. Comoving coordinates are no longer
a particularly good choice within a collapsed halo.

In our universe a large halo might have mass $10^{15} \Msun$.  Today's
$\rho_m \approx 3.3 \times 10^{10} \frac{\Msun}{Mpc^3}$ gives a
corresponding comoving volume of about $3 \times 10^4 \mathrm{Mpc}^3$,
which is nearly a factor of a billion smaller than the maximum
comoving size.  Any late-time entropy source must therefore compensate
for effectively having a causal diamond
  3-volume approximately  $10^{-9}$ of that 
during peak stellar entropy production. Whether this is possible
depends upon details such as the lengths of time available and the scale of
entropy production. The most dramatic example would be the inclusion
of Hawking radiation from a black hole. Release of a $10^7 \Msun$
black hole's $10^{91}$ entropy as Hawking radiation would completely
swamp entropy produced by stars in the first $10^{10}$ years of cosmic
evolution.  The time scale is enormous: $\mathrm{log}_{10}(\tau_{BH})
= 83 + 3\mathrm{log}_{10}[M_{BH}/10^6 \Msun]$, or about $10^{86}$
years in this case, but we cannot ignore the situation out of hand as
any worldline which tracks matter has a high chance of ultimately
ending up near (or even in) a black hole.
\section{Why we care about particular world lines}
In the formulation of Bousso {\it et al.} \cite{Bousso} the universe is
considered to be exactly FRW homogeneous and isotropic, which makes a
distinction between comoving volume and mass unnecessary. Indeed over
the size of the causal diamond this assumption is quite accurate at
the beginning of our universe and well through the current time, as
the universe is homogeneous well below scales approaching the Hubble
length or the current size of the causal diamond.  As mentioned
above it is in the future that differences among causal diamonds may
arise.

Because of the assumption of homogeneity in \cite{Bousso}, a
particular set of cosmological parameters resulted in a unique,
representative causal diamond.  Thus the probability is given by
$$\frac{dp}{d\rholambda} \propto
w(\rholambda)\frac{dp}{dN}\frac{dN}{d\rholambda} $$ where
$dN/d\rholambda$ represents the density of vacua per value of
  $\Lambda$.  We may take $dN/d\rholambda$ to be flat if the
landscape has values 
  spaced tightly in the region of interest, and if 0 is not a special
  value. With these assumptions, the spacings of vacua can be assumed
  to be uniform for $\Lambda$ near $10^{-123}$. The quantity
  $dp/dN$ is the 
  term representing the theory's prior probability for $\Lambda$.
  Following previous work, we assume prior probability is flat; in
  other words, the background theory is indifferent to vacua,
  choosing among them with equal probability. 

Critically, in \cite{Bousso}, the weighting $w(\rholambda)$ is the
weight of a single representative causal diamond with cosmological
constant density $\rholambda$. If a particular set of cosmological
parameters does not yield a single causal diamond, we must replace our
single calculation of $w(\rholambda)$ with a probability distribution
$$w(\rholambda) = \int_{\lambda}w(\rholambda,\lambda)\mathrm{d}\lambda$$
where the integral over $\lambda$ is one over all possible world lines
(and hence causal diamonds) given a particular set of cosmological
parameters from our background theory. 

This discussion may seem counter to the spirit of the causal diamond
approach in \cite{Bousso}.  Yet unless our background theory is itself
phrased in terms of causal diamonds, we cannot skip smoothly from a
distribution of cosmological parameters to a distribution of results 
for causal diamonds. Our prior distribution of $\Lambda$ or the
spacing of vacua is phrased in terms of cosmological parameters, not
particular world lines.  Assuming perfect homogeneity simply means
taking the weight function $w(\rholambda, \lambda)$ to be proportional
to a delta function peaked at a particular world line $\lambda_0$
that is ``typical'' of a perfect FRW universe.  Given the tremendous
variety of world lines for any structure-forming cosmology, this
assumption seems unrealistic: an extreme counterexample would be a world line
that runs directly into a black hole horizon at an early
era. Nonetheless it remains to be seen whether considering an ensemble
of world lines for a cosmology rather than a single one makes a difference in
predictions for $\rholambda$.

In order to calculate the entropy production probability distribution
over an ensemble of world lines ${\lambda}$, we need to describe how
the density of a bundle of world lines behaves over time relative to the
coordinates in which we wish to measure entropy production. We argue
that for an inhomogeneous universe there are multiple ways to
parametrize these world lines and that the choice of parametrization
  directly affects the results of CEP calculation.

It should be noted that even in the case of a perfectly homogeneous
FRW universe not all world lines (and hence causal diamonds) are
created identically, as one could imagine arbitrary boosts or even
accelerated paths relative to a comoving observer.  Even with modest
boosts, observers on these paths would have a different experience of
the universe owing for example to a strong CMB dipole. Since the group
of boosts is not compact, one might expect a ``typical'' boost
to be arbitrarily far from the comoving rest frame, with
correspondingly anisotropic physics. Given a homogeneous, isotropic
universe, the preservation of symmetry afforded by the choice of a
comoving observer seems an enticing motivation for picking a comoving
causal diamond. But it must be emphasized that this is indeed a
choice, and any appeal that comoving coordinates are natural in the
sense that they follow typical matter distributions (and perhaps thus
observers) has implications for the inhomogeneous case.

When we move to an inhomogeneous universe we cannot even appeal to a
notion of preserving symmetry.  For the purposes of simplification we
will leave out accelerated world lines and describe our collection of
world lines as a congruence of timelike geodesics, with each spacetime
point lying on a single geodesic. One can construct such a congruence
by specifying a spacelike slice and examining geodesics orthogonal to
this slice. Different slices, however, typically result in different
inherited parameterizations for the world lines. We will describe two
such choices in what follows, but there are of course many others.

For a slice picked at a constant cosmic time in the very homogeneous
early stages of a universe like ours, there is a natural parametrization: our
entropy production can be measured on a per-mass or, equivalently,
comoving coordinate basis, and so we can simply imagine a grid of
world lines piercing each spacelike surface with constant cosmic
time. In the homogeneous limit for comoving coordinates this grid
simply remains fixed in time, yielding a fixed world line density, and
corresponding to the simple choice made in Bousso {\it et
al}. \cite{Bousso}

The generalization to a more realistic, slightly inhomogeneous universe requires one to make further choices. A starting point is to imagine placing test particles in
a fixed, constant spatial density at an early cosmic time, and
watching the particles trace out geodesics as the universe evolves.
Of course, our universe seems to have performed this very experiment,
and as $\Lambda$ dominates we have a picture of most matter eventually
residing within isolated gravitationally bound halos, with exponentially emptying space in
between. In this picture, at late times the spatial density distribution
of geodesics parallels that of matter itself, so at least roughly, a
probability distribution for entropy production over world lines would
be equivalent to integration over the matter distribution. This ``early comoving'' choice could also be motivated by
a combination of constraints on the initial distribution of world
lines and enough inflation to turn the initial distribution into a
comoving one.

There are other choices that yield dramatically different answers,
however. If we choose a slice at late cosmic time and parametrize
world lines to have a constant density in physical coordinates, the
vast majority of world lines at late times will be located in nearly
empty regions with almost no entropy production.  When we trace back
the world lines to the beginning of the universe, they will not be
homogeneously distributed relative to matter, but for the purposes of
calculating entropy increase at early times there is no significant
difference since the entropy production itself is homogeneous in
space.

On the other hand, with this second choice, any entropy production at
late times will be exponentially suppressed by the rarity of world
lines that are located near matter, and so given this choice it is
justifiable to discard late-time entropy sources. It is important to
observe, however, that the second choice seems at best no better motivated
than the first, and indeed that one could imagine many other
intermediate choices for parameterizing world lines. For the remainder
of the paper we treat this choice as an open question, and will
estimate the effect of late-time entropy production where it seems to matter:
that is, under the ``early comoving'' assumption that typical world lines follow matter
distribution from an early time.  Therefore we begin by asking what
astrophysical processes may produce substantial entropy well into the
future.  As we will discover in Sec. V.E, carefully considering scenarios in the very far future provokes additional questions about the distribution of world lines.

\section{Long-term entropy production}

\subsection{Black holes}
Black holes contain much more entropy than all other astrophysical
sources. In \cite{Bousso}, black hole horizon entropy as well as that
associated with the formation of a de Sitter horizon were explicitly
excluded from the tally of entropy increase. Maor et al. \cite{Maor}
raise the possibility that gravitons produced during black hole
mergers could by themselves exceed stellar entropy increase. But even
if one does not count a significant early-time increase in entropy
from black holes, on the very long time scale of black hole
evaporation, this entropy increase can no longer be avoided. Hawking
radiation returns entropy to the matter sector, and it will typically
dominate the early-time stellar entropy production as estimated in
Sec. III.

\subsection{Stellar entropy}
Low-mass white dwarfs may continue burning for as long as $10^{13}$
years.  Moreover, even though star formation is already dropping
dramatically in our universe due to depletion of cool gas, some small
but finite star formation rate will likely exist far into the future
owing to collisions among substellar masses and white dwarfs.
Further, one might wonder about the time behavior of star formation in
universes with very different values of $\Lambda$.

Can stars in a collapsed region far into the future ever exceed the
$10^{86}$ entropy produced by the stars in the first $10^{10}$ years?  We
can calculate an 
upper bound by simply imagining all baryons within a halo are
converted into stars and burned. Consider a massive halo ($10^{15}
\Msun$).  Baryons make up about 1/6 of the matter content, or $1.6
\times 10^{14} \Msun = \frac{3 \times 10^{44} kg}{2 \times 10^{-27}
kg/baryon} \approx 1.5 \times 10^{71}$ baryons, or perhaps $10^{71}$
hydrogen atoms.

Each instance of fusion releases about 7 MeV per baryon.  At a
temperature of 20 meV for dust-reprocessing, that is about $3 \times
10^8$ entropy per dust-processed baryon.  Even if over very long times
100\% of baryons are burned to hydrogen, and half are reprocessed by
dust (an overestimate as dust is depleted over time), that allows only
$ \approx 10^{79}$ entropy, 7 orders of magnitude less than is produced
by stellar entropy $\Delta S_0$ up to $10^{10}$ years.  It would seem
that for the observed cosmological parameters future stellar entropy
production can not compete with that in the past.

Varying the cosmological constant affects the estimate in two ways:
increasing $\Lambda$ leads to earlier vacuum domination and a smaller
value of $\Delta S_0$. However, it simultaneously leads to a smaller
typical halo size as discussed later. Eventually large $\Lambda$ will
lead to a severe drop in star formation rates at both early and later
times. Similarly, small values of $\Lambda$ will push vacuum
domination later and later, eventually leaving less stellar entropy to
be produced in the vacuum-dominated era. Thus it does not appear that
stellar entropy in late eras is a strong competitor to $\Delta S_0$,
even when the cosmological constant is varied.

\subsection{Dark matter annihilation}
To compete with $\Delta S_0$ we need approximately $10^{15}$ entropy
per baryon.  With the possible exception of Hawking radiation, this
appears to be a tall order.  We
need a process with a combination of high energy released, low
effective temperature, and near universal occurrence.  One possibility
is annihilations of dark matter.  Dark matter masses perhaps 6 times
baryonic matter, so the total available energy is $\approx$ 6 GeV per
baryon.  Cline et al. \cite{Cline} chose a low-mass dark matter model in which decays could compete even at early times with stellar entropy, and used the CEP to make predictions for this model.  Because our paper focuses on possible effects on the CEP for entropy sources at late times, we will instead consider a simple model of an annihilating weakly interactive massive particle (WIMP) with approximately weak scale mass and estimate the entropy production over very long time scales.  With such a high mass, the typical handful of
WIMP annihilation products by themselves cannot produce anywhere near
enough entropy.  So the interesting case is if the annihilation
happens in a low-temperature context so that many low-energy products
(typically photons) can be produced by a single annihilation.

Adams {\it et al.} \cite{Adams} explore WIMP capture by white dwarfs.  Over
the long term, white dwarfs make up the bulk of collapsed stellar objects, and
they have densities great enough to capture massive WIMPs over time.
Due to dark matter (DM) annihilations the dwarfs have a very extended period of low
luminosity and low temperature.  Adams {\it et al.} give typical T $\approx$ 63 K, or
about 5 meV for DM annihilations, which with 6 GeV/baryon energy gives
only $10^{12}$ entropy from annihilating all DM.  
\subsection{Proton decay}
One can also ask about proton decay within white dwarfs (88\% of final
stellar mass).  For a typical grand unified theory decay such as $p \to e^+ + \pi^0$, 1
GeV per nucleon is ultimately released.  Typically about 1/3 is lost
to neutrinos which freely stream out of even white dwarfs rather than
thermalizing.  Thus we need a temperature of T $\approx 10^{-6}$ ev or
about $10^{-2}$ K.  For proton decay in white dwarfs, T $\approx$ .06
K with proton decay lifetime $\Gamma = 10^{37}$ years.  Using the same
bounds on proton decay as Adams {\it et al.}\cite{Adams}, $32 < \log\Gamma < 41$, but since $T^4
\propto \Gamma e^{-\Gamma t}$, we can only push that temperature down
another order of magnitude with the simplest proton decay models.  But
a proton decay mechanism originating from a higher order operator
could produce much longer lifetimes and correspondingly lower
temperatures, perhaps allowing this process to compete with early stellar
evolution.
\subsection{Dynamical effects}
Given the approximations involved, either proton decay or WIMP
annihilation might be considered reasonable competitors to stellar
entropy production $\Delta S_0$ in the matter-dominated era.  In order to
calculate the maximum entropy for each we have simply given each
process a maximal value assuming complete conversion of a certain
large halo.  But halo masses themselves may not be stable on the time
scales considered ($\tau \approx 10^{24}$ years for WIMP annihilation
and $\approx 10^{37}$ years for proton decay).  There are two competing
dynamic processes within halos over the very long term (\cite{Adams}).
Interactions between stars lead to dynamic relaxation and ejection of
individual stars on a time scale of $\tau_{evap} \approx
100\tau_{relax} \approx 100\frac{R}{v} \frac{N}{12 \ln(\frac{N}{2})}
\approx 10^{19}-10^{20}$ years for typical galactic radius $R$, random
velocity $v$, and number of stars $N$. At the same time, gravitational
radiation should cause orbits to decay and eventually drop matter into
a central black hole, on a time scale of $\approx 10^{24}$ years.
Adams {\it et al.} estimate perhaps 1-10\% of matter remains bound to the
central black hole while the remainder is lost from the galaxy.

Matter ejected from the gravitational bounds of a galaxy will in
general be lost from the de Sitter horizon as well. Taking the
point of view of a world line following an example white dwarf ejected
this way, within a few Hubble times the former host halo will have
redshifted beyond the horizon and the only continuing source of
entropy increase within an observer's horizon and causal diamond would
be that produced from the single white dwarf star.  Even the complete
proton decay of such a star would produce a completely negligible
amount of entropy compared to $\Delta S_0$ given the small matter
content within the horizon.  On the other hand, we may still wonder
about a single large black hole ejected in this fashion, since Hawking
radiation over extremely long times could compete with early entropy.

If this picture of dynamical effects is correct, for a world line near
the leftover central black hole in a halo, of the processes considered
again it is only Hawking radiation that could compete with $\Delta
S_0$, as on time scales much shorter than proton decay, essentially
all matter will have either been ejected from the halo or have already
collapsed into the central black hole.  WIMP annihilation within white
dwarfs has a time scale of $\approx 10^{25}$ years, so the story is
relatively similar:  white dwarfs experiencing this process will typically be 
isolated and the resulting entropy gain will not be within the causal diamond 
of the bulk of remaining matter. 

Dynamical effects may also have important implications for
counting entropy from Hawking radiation. We intentionally made
the choice to parametrize world lines so that they essentially
followed typical paths of matter. We claimed that this choice was in essence
arbitrary, if straightforward. On time scales well before black hole
evaporation, there are only two common fates for matter:
either it is within a black hole, or part of a small amount of
matter with no black holes within the de Sitter horizon. The question
then becomes how literally to accept this picture when talking about
idealized world lines rather than actual matter particles.

If we imagine our geodesics to be the paths of perfect test particles
in a zero-mass limit, it is easy to imagine that these orbits are
stable over very long periods.  The time estimates earlier quoted for
dynamical ejection or decay by gravitational radiation were for
ordinary massive objects with typical virial velocities.  In
comparison, for the simple case of a perturbations to a geodesic
orbiting a black hole (due to e.g. Hawking radiation, or classical
estimates from the changing mass of the black hole), there is
ultimately just one time scale: that of the black hole evaporation
itself. Therefore such world lines would be exposed to the bulk of
black hole evaporation entropy, and if they are common they could
easily ruin the CEP predictions, as discussed in the next section.

We might take a different view: that we are really interested in
typical world lines associated with matter, and that we should thus
turn away from perfectly idealized geodesics in favor of the attractor
behavior for matter far in the future.  In this case it is quite
common for world lines themselves to intersect black holes, but
it is not common for world lines to stay within a Hubble radius of a
black hole long enough to observe black hole evaporation. Classical
world lines may end at the singularity of a black hole, but our
approach of ignoring horizon entropy is not nearly so obvious once the
horizon itself is crossed. In this picture the CEP may be safe from
the need to count Hawking radiation, but the details are far from
immediate.

It should be noted that the differences between the above approaches
are very subtle in practice except perhaps at times long in the
future, and that both are variants of the ``early comoving'' choice
that we have already explicitly made. Subtle changes in
parametrization clearly can yield very different ideas of what
constitutes a typical world line, and we next make the argument that if
there is a significant late-time entropy source in our
parametrization, it can drastically affect the CEP prediction for a
cosmological parameter.

\section{Effects on prediction for $\rholambda$}
All of our late-time effects are at approximately fixed mass within a
horizon. Assuming we are examining an astrophysical process which is
independent of halo scale (which is certainly true for proton decay
itself, but should be considered a simplification for black holes and
white dwarf processes, since larger halos may have different
astrophysics), the only determinant of entropy production is the mass
of the halo.  Diamonds containing different masses would also have slightly different volumes owing to the different metric compared to an empty de Sitter diamond, but entropy production at late times takes place within compact objects, and does not depend on the volume of the causal diamond.  Thus we describe our entropy weight as a
function of the mass of the halo, $w(\rholambda, \lambda) \rightarrow
w(\rholambda, M_{halo})$.  If we assume matter fairly traces
worldlines (the ``early comoving'' assumption), then we can take
advantage of the Press-Schechter (P-S) mass function\cite{PS} to estimate
the probability for each world line (and thus causal diamond) to be
within a halo of mass M and hence to have a weight $w(\rholambda,M)$.

\begin{figure}[h]
  \includegraphics[height=.3\textheight]{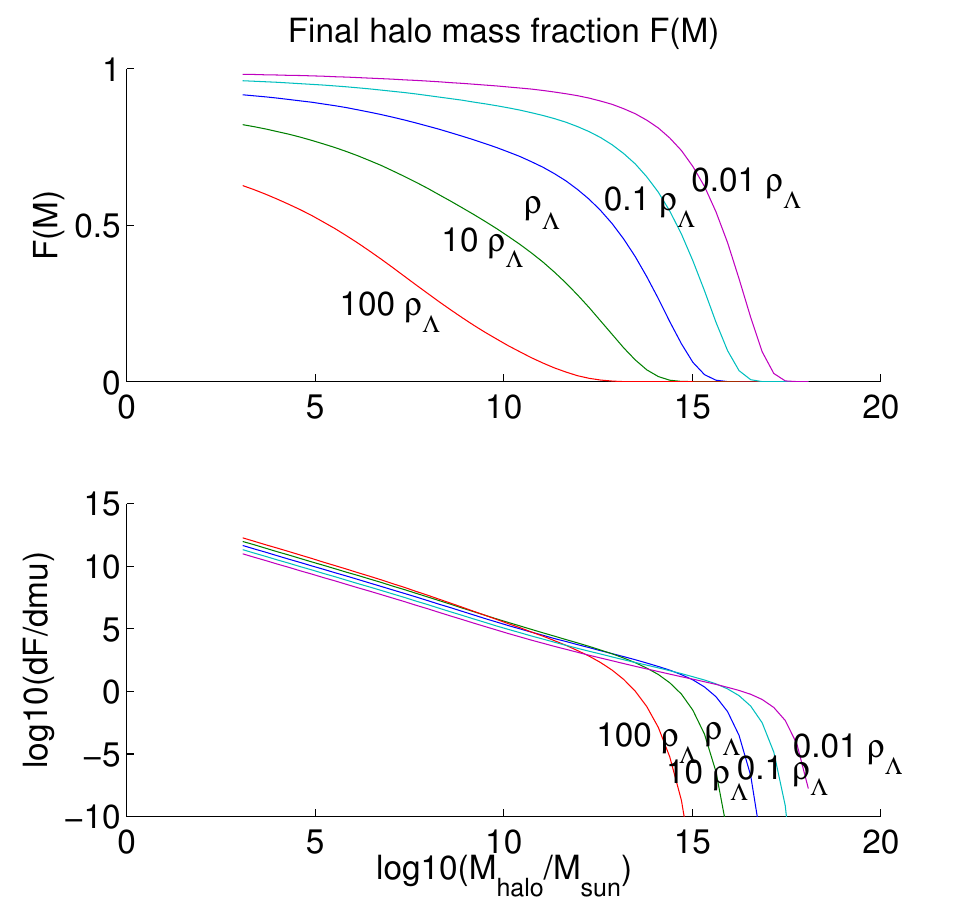}
 \caption{Top: Fraction of total mass at late times in halos smaller than mass M,
 plotted for a range of cosmological constants relative to the
 observed value $\rholambda$. Bottom: the late-time differential Press-Schechter
 halo fraction f(M). As $\Lambda$ increases, the halo fraction shifts towards smaller masses.  }
 \label{halofrac}
\end{figure}

The P-S mass function $F(M, \Lambda) = \mathrm{erfc}[\frac{\delta_c}{\sqrt{2} \sigma(M,\Lambda)}]$ gives the fraction of total mass contained in haloes of mass larger than $M$. 
In a $\Lambda$-dominated universe, these mass fractions approach
a fixed value on the timescale of $t_\Lambda \approx$ 16.7 billion
years for our universe, as the cosmological constant freezes structure
formation\cite{Tegmark}.  The final halo mass fraction and
differential P-S function are given in Fig. \ref{halofrac}, assuming the functional form for $\sigma$ given in \cite{Tegmark}.

For calculations of total entropy production at late times only the mass of the halo should be relevant.  For many of the potential sources mentioned, such as particle decays, we can approximate the entropy production as simply proportional to the halo mass.  For Hawking radiation from black holes, $S \propto M^2$, so if the black hole mass is proportional to the halo mass, the entropy production will be proportional to the square of the halo mass.  These correspond to different choices for the weight $w(M)$, while the differential P-S mass function $-\dFdM$ will give the probability distribution for halo masses as a function of $\Lambda$.  
Using a general power-law weighting $\Delta S \propto M^n$ for halo mass M, we have:

$$w(\rholambda) \equiv \int w(\rholambda,\lambda) \mathrm{d}\lambda$$
$$= \int_0^{\infty} w(\rholambda,M) p(M) \mathrm{d}M$$
$$= -\int_0^{\infty}  M^n \dFdM(M,\Lambda) \mathrm{d}M $$
$$= \int_0^{\infty} M^{n-1} \mathrm{erfc}[\frac{\delta_c}{\sqrt{2} \sigma(M,\Lambda)}]\mathrm{d}M$$

\begin{figure}[h]
  \includegraphics[height=.5\textheight]{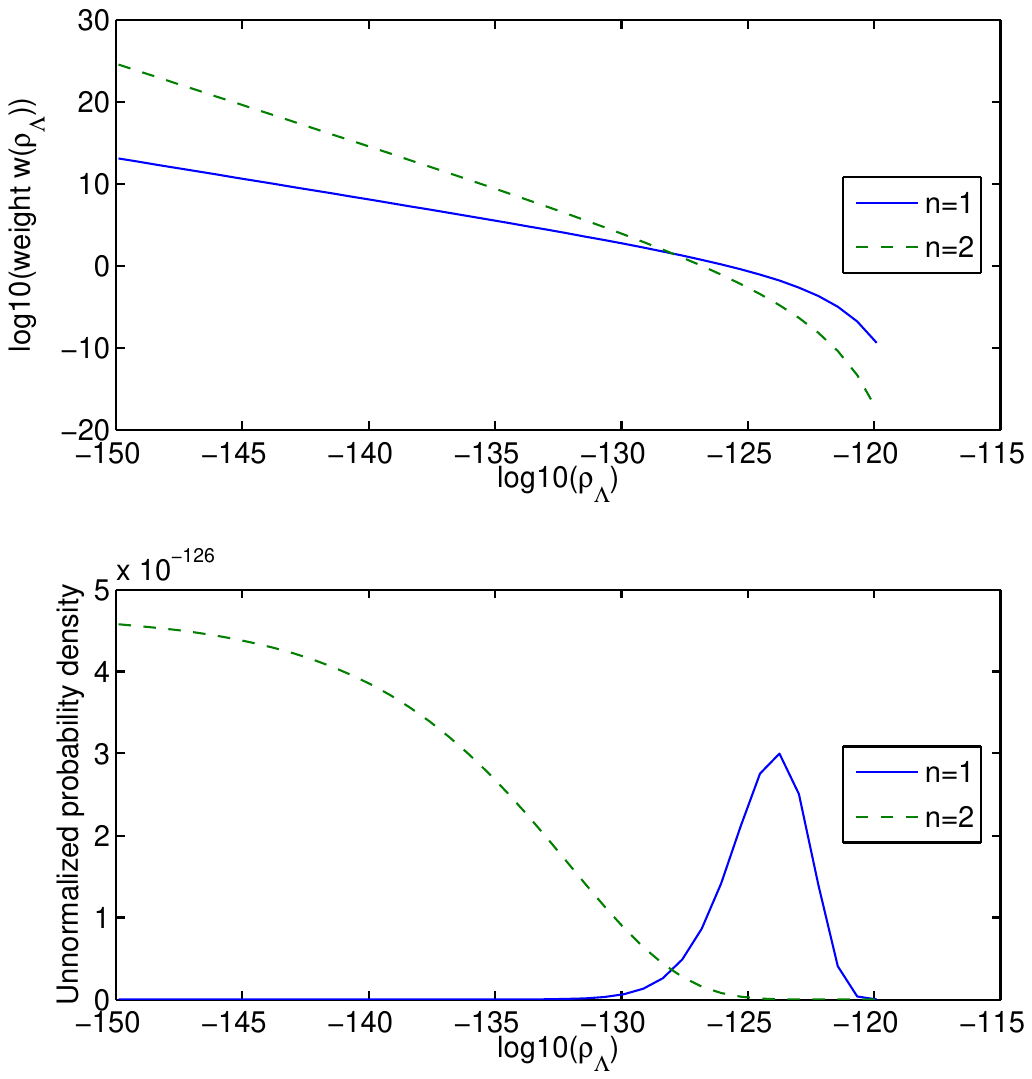}
 \caption{Top: Unnormalized weight $w(\rho_\Lambda)\propto \int M f(M,\infty) \mathrm{d}M$
 assigned to a constant entropy production after $\Lambda$ domination, plotted for $\Delta S \propto M$(solid) and $M^2$ (dashed). Bottom: corresponding probability densities, $\frac{\mathrm{d}P}{\mathrm{dlog}(\rholambda)} \propto w(\rholambda)\rholambda$. The $n=1$ case provides a peaked prediction for $\mathrm{log}(\rholambda)$, while the $n=2$ case does not.}
 
 \label{futureweight}
\end{figure}

In the case of $n=1$, the weight given to a given value of $\Lambda$ is simply 
proportional to the mass of the halo a typical piece of matter finds itself within. One expects this typical halo mass to decrease for larger $\Lambda$.  Indeed, from Fig. \ref{halofrac}, we can see that for $\Lambda$
within a few orders of magnitude of the observed value, $\dFdM(M,t)$
scales approximately inversely in mass with increasing $\rholambda$.  Below and in Fig. \ref{futureweight} we analyze in the important small-$\Lambda$ limit of this scaling behavior.

In the late universe $\delta_c$ reaches an asymptotic value of 1.63, while the typical fluctuation on a mass scale M may be factorized at late times as \cite{Tegmark} 
$$ \sigma(M,\Lambda) \propto \frac{s(M)}{\Lambda^{1/3}}$$
where $s(M)$ is a function that, for fixed matter-to-light ratio, depends solely on the halo mass.  $s(M)$ grows logarithmically towards small masses while decreasing as $M^{-\frac{2}{3}}$ for large masses.  The break in the power-law occurs at the horizon mass at matter-energy equality: smaller mass scales entered the horizon in a radiation-dominated universe and had a period of suppressed growth.  We will be interested in estimating the asymptotic $\Lambda$ dependence of the integral for small values of $\Lambda$, where typical haloes are far larger than this critical mass scale and the integral is dominated by the large $M$ power-law behavior.  So in this limit:

$$w(\rholambda) \propto \int_0^{\infty}  M^{n-1} \mathrm{erfc}[C \Lambda^{1/3} M^{2/3})] \mathrm{d}M $$
for some constant $C$
$$ \propto \Lambda^{-n/2}\int_0^{\infty}  x^{\frac{3n}{2}-1} \mathrm{erfc}[x]\mathrm{d}x$$
with a suitable change of variables.  


For the $n=2$ case associated with counting late-time entropy from black holes, $w(\rholambda) \propto \rholambda^{-1}$.  This replicates (for entirely different reasons) a result from Sec. II, where a flat prior on $\rholambda$ combined with a constant-rate entropy source and a causal diamond volume $V \propto \rholambda^{-1}$ leads to a flat predicted distribution for
$\rholambda$ in logarithmic space:

$$\frac{\mathrm{d}P}{\mathrm{dlog}(\rholambda)} \propto w(\rholambda)\rholambda \propto \text{const}$$

A flat probability distribution in logarithmic space provides no explanation for the observed order of magnitude of $\Lambda$; indeed, it is a prediction that $\Lambda = 0$, or else the smallest discrete value allowed by a theory.  Higher values of $n$ will simply lead to more sharply peaked predictions of $\Lambda = 0$.

For the $n=1$ case, we recover a peak in the prediction for $\rholambda$ that is very similar to the results in ref. \cite{Bousso}.  It is important to note that the peak is driven by the optimization of late-time entropy sources and represents the result of a competition between increased typical halo size associated with smaller values of $log(\Lambda)$, and the corresponding rarity of those cosmologies.  In the original work it was the increased amount of stellar entropy produced before $\Lambda$ domination, rather than typical halo size, that pushed for smaller values of $\Lambda$.  Although these effects are not entirely unrelated, the fragility of the correspondence between the early- and late-time calculations is shown by the rather different behavior under other assumptions of the scaling between halo mass and entropy.


We have focused on calculations of late-time entropy production, $\Delta S_{late}$, but the overall weighting for a cosmology will be $\Delta S_0 + \Delta
S_{late}$.  In the case where the late-time entropy increase dominates, the flat
prior in $\rholambda$ may be transformed to a flat distribution for
$\log(\rholambda)$.  While this result addresses the cosmological
constant problem to some degree (as we need only explain the smallness
of log $\Lambda$), compared with earlier work we have lost the peak in
the probability distribution associated with a prediction of the
actual value of $\Lambda$.  Unless we can definitively rule out significant
late-time entropy sources with $\Delta S \propto M^n, n\ge 2$, such a result would undermine
some of the success of the CEP.  Nonetheless the CEP
still benefits to an extent from the suppression of structure formation for
large values of $\rho_\Lambda$, which shows up in the weighting
dropping quickly for values of $\rholambda$ significantly larger than the observed value (as can be seen in Fig. \ref{futureweight}).

\section{Conclusions}

Standard treatments of the causal entropic principle consider a
one-to-one mapping between cosmological parameters and causal
diamonds. The inhomogeneity of a realistic universe introduces
additional complexity because different observers can experience very
different causal diamonds, even with the same cosmological parameters.
One must have some method of picking a typical causal diamond, or of
characterizing an ensemble of causal diamonds for a given
cosmology. We have shown that with one reasonable choice of
parameterizations for the ensemble of causal diamonds, we are forced to
consider very slow entropy sources in the far future.  Dynamical
effects on the typical halo over long times may 
prevent these slow entropy sources from being important contributors
to the overall measure, but it is easy to imagine particular
parameterizations where this is not the case.  The entropy associated
with black hole evaporation or certain models of particle decay could
then ruin CEP predictions for the value of the cosmological constant.

It should also be noted that in a universe with enough inhomogeneity
and with smaller causal diamond sizes, the effect of the inhomogeneity
would be pushed to earlier time scales and we would need to worry
about the clumping of stellar entropy production itself rather than merely
late-time events. An example would be a universe with much larger
$\Lambda$ and also much larger initial fluctuations.

There are methods to parametrize causal diamonds that seem to
avoid the late-time entropy production issue discussed here for
universes similar to ours. But this ambiguity seems to point at least
to an incompleteness in the CEP as currently formulated. One could of
course simply make a felicitous choice of parametrizations and add it
to the CEP. But for a wide range of cosmological parameters it may be
still be difficult to be sure of capturing a typical causal
diamond in this fashion. The reliance on entropy associated with a
single causal diamond makes this issue much more difficult than it would be
for (e.g.) a per-baryon measure, and in that sense is a CEP-specific
issue. And it is one that must be addressed to be confident of CEP predictions for
nonidealized cosmologies.

\begin{acknowledgments}
We wish to thank S. Carlip, R. Bousso, B. Freivogel, D.
Martin and B. Bozek for many fruitful discussions. This work was
supported by the DOE under Grant No. DE-FG03-91ER40674.

\end{acknowledgments}


\bibliography{entropy}
\bibliographystyle{apsrev}
\end{document}